# Thermal decomposition of n-dodecane: experiments and kinetic modeling

Olivier Herbinet, Paul-Marie Marquaire, Frédérique Battin-Leclerc, René Fournet,

Département de Chimie Physique des Réactions, UMR 7630 CNRS - INPL,

1 rue Grandville, BP 20451, 54001 NANCY Cedex, France

**Abstract**

The thermal decomposition of n-dodecane, a component of some jet fuels, has been studied in a jet-stirred reactor at temperatures from 793 to 1093 K, for residence times between 1 and 5 s and at atmospheric pressure. Thermal decomposition of hydrocarbon fuel prior the entrance in the combustion chamber is an envisaged way to cool the wall of hypersonic vehicles. The products of the reaction are mainly hydrogen, methane, ethane, 1,3-butadiene and 1-alkenes from ethylene to 1-undecene. For higher temperatures and residence times acetylene, allene, propyne, cyclopentene, 1,3-cyclopentadiene and aromatic compounds from benzene to pyrene through naphthalene have also been observed. A previous detailed kinetic model of the thermal decomposition of n-dodecane generated using EXGAS software has been improved and completed by a sub-mechanism explaining the formation and the consumption of aromatic compounds.

# 1. Introduction

One of the main issue of hypersonic flight is the thermal management of the vehicle and more especially of the engine [1,2]. Studies are currently conducted in order to develop an active cooling system, called regenerative cooling, that consists in using the fuel (typically a blend of heavy hydrocarbons) heat sink capability due to the endothermicity of its thermal decomposition [2-5]. Because of the large heat load found in a hypersonic vehicle, engine temperature is expected to increase well above 800 K leading in the thermal decomposition of the fuel circulating at their walls before the injection in the combustion chamber. A consequence of the decomposition of the heavy hydrocarbon fuel is the production of lighter species and then of aromatic and poly-aromatic compounds. In order to quantify the heat transfer through the walls of the engine and the composition of the fuel entering the combustion chamber, a detailed kinetic model of the thermal decomposition of the fuel is required.

Jet fuels are composed of hundreds of aliphatic and aromatic hydrocarbon compounds of which major components are normal alkanes, branched alkanes, cycloalkanes, aromatics and alkenes. Surrogate mixtures have been developed in order to reduce the size of pyrolysis and oxidation models of jet fuels. N-dodecane, a major component of some jet fuels (kerosene, NORPAR12, JP7, JP8 [4]) is one of the surrogates usually chosen for normal alkanes [6]. A very limited work was published about n-dodecane pyrolysis. Zhou et al. studied the thermal decomposition of n-dodecane in an unpacked flow reactor at atmospheric pressure and at temperatures from 623 to 893 K [7]. Yoon et al. performed the pyrolysis of pure n-dodecane in a batch reactor at a pressure of 10 bars and at temperatures from 673 to 523K [8,9]. More recently a new study of the thermal decomposition of n-dodecane allowed us to propose a detailed kinetic model made of 1175 reactions involving 175 molecular and radical species [10]. This model was mainly validated from experimental results obtained in a plug flow reactor at higher temperatures (950-1050 K) than the previous papers [7-9] and residence times ranging from 0.05 to 0.2 s. In order to get experimental data in a different type of reactor and for a broader range of experimental conditions the new study presented here has been performed using a jet-stirred reactor, which is a well suited apparatus for gas phase kinetic study [11-14].

## 2. Experimental method

The experimental apparatus (Figure 1) which was developed for the study of the thermal decomposition of liquid hydrocarbons involves a new device used for the liquid hydrocarbon feed and evaporation control and a quartz jet-stirred reactor (JSR).

**Figure 1**

Liquid n-dodecane (reactant) is contained in a glass vessel pressurized with nitrogen. Before performing a series of experiments, oxygen traces are removed from the liquid hydrocarbon through nitrogen bubbling and vacuum pumping. Liquid hydrocarbon flow rate between 1 and 50 $g.hr^{-1}$ is controlled, mixed to the carrier gas (helium) and evaporated in a Controlled Mixer and Evaporator (CEM) provided by "Bronkhorst". This apparatus is composed of a liquid mass flow controller followed by a mixing chamber and a single pass heat exchanger. Carrier gas flow rate is controlled by a "3 $nL.min^{-1}$ Tylan – RDM280" gas mass flow controller set upstream the CEM. Molar composition of the gas at the outlet of the CEM is 2% hydrocarbon and 98% helium. Working with high dilution presents two advantages: the variation of molar flow rate due to the reaction is very weak and can be neglected and strong temperature gradients inside the reactor due to endothermic reactions are avoided.

The thermal decomposition of the fuel was performed in a quartz jet-stirred reactor that was developed by Matras and Villermaux [15]. This type of reactor (Figure 2) which was already used for numerous gas phase kinetic studies [11-14] is a spherical reactor in which diluted reactant enters through an injection cross located in its center and that can be considered as well stirred for a residence time (τ) between 0.5 and 5 s [15,16]. The diameter of the reactor is about 50 mm and its volume is about 90 cm$^3$. The stirring of the whole reactor volume is achieved by the mean of four turbulent gas jets directed in different directions and produced by the four nozzles of the injection cross in the center of the reactor. Inside diameter of the nozzles is about 0.3 mm. The reactor is preceded by a quartz annular preheating zone in which the temperature of the vapor mixture is increased up to the reactor temperature before entering inside. The annular preheater is made of two concentric tubes, the inter-annular distance of which is about 0.5 mm. Gas mixture residence time inside the annular preheater is very short compared to its residence time inside the reactor.

Both spherical reactor and annular preheating zone are heated by the mean of "Thermocoax" heating resistances rolled up around their wall. Temperature control of each part is made by "Eurotherm 3216" controllers and type K thermocouples. Reaction temperature measurement is made by means of a thermocouple which is located inside the intra-annular space of the preheating zone and the extremity of which is on the level of the injection cross.

**Figure 2**

Pressure inside the JSR is close to atmospheric pressure (about 106 kPa). It is manually controlled thanks to a control valve placed after the liquid hydrocarbon trap.

32 products of n-dodecane thermal decomposition were identified and quantified by gas chromatography and can be classified in two groups. First, light species (on-line analysis) which are hydrogen and $C_1$-$C_5$ hydrocarbons: methane, acetylene, ethylene, ethane, allene, propyne, propene, 1,3-butadiene, 1-butene and 1-pentene. Second, $C_5$-$C_{16}$ species: 1,3-cyclopentadiene, cyclopentene, benzene, 1-hexene, toluene, 1-heptene, phenyl-ethylene, 1-octene, indene, 1-nonene, naphthalene, 1-decene, 1-undecene, 1- and 2-methyl-naphtalene, biphenyl, acenaphthalene, phenanthrene, anthracene, fluoranthene and pyrene.

Light species samples are taken at the outlet of the reactor and are analyzed on-line by two gas chromatographs set in parallel. The first gas chromatograph is a "Hewlett Packard 5890" equipped with a thermal conductivity detector (TCD) for hydrogen detection, a flame ionization detector (FID) for hydrocarbons detection and a carbosphere packed column. A first analysis is made at an oven temperature of 473 K for methane and $C_2$-hydrocarbons detection. A second analysis at an oven temperature of 303 K is necessary to separate hydrogen peak from helium peak (experiment carrier gas). Argon was chosen as column carrier gas and TCD reference gas in order to detect hydrogen with a better sensibility. The second gas chromatograph used for $C_3$ and $C_4$-hydrocarbons is a "Schimadzu 14A" fitted with a FID and a Haysep-D packed column with nitrogen as carrier gas. Oven temperature profile is: 313 K held 30 min, rate 1 K.min$^{-1}$, 473 K held 30 min.

Heavy species (Figure 3) are accumulated in a glass trap quenched in liquid nitrogen (about 77 K) during time typically between 5 and 15 min according to reactant feed. Then the glass trap is disconnected from the reactor and products are dissolved in a solvent (acetone). A known amount of internal standard (n-octane) is added to the mixture which is injected by an auto-sampler in an "Agilent 6850" gas chromatograph equipped with a FID and a HP-1 capillary column with helium as carrier gas. Oven temperature is: 313 K held 30 min, rate 5 K.min$^{-1}$, 453 K held 62 min. Heavy species were identified by a gas chromatograph-mass spectroscopy (GC-MS) system working in the same conditions as the "Agilent 6850" gas chromatograph.

**Figure 3**

## 3. Experimental results

Experimental results have been obtained at temperatures from 793 K to 1073 K, residence times ranging from 1 to 5 s, and with a reactant mole fraction at the inlet of the reactor of 2%. On the following graphs (Figure 4 to , Figure 6), symbols correspond to experiments and curves to modeling.

Figure 4 represents the evolution of the conversion of n-dodecane vs. residence time (Figure 4a) and vs. temperature for a residence time of 1 s (Figure 4b). It shows that the conversion strongly increases with temperature between 873 K and 1023 K.

**Figure 4**

Figure 5 and Figure 6 present the evolution of the mole fractions of the products of the reaction vs. residence time (Figure 5) and vs. temperature (Figure 6). The major products are hydrogen and ethylene. For most of the pyrolysis products, an increase of the residence time or of the temperature leads to an increase of the mole fraction (e.g. hydrogen, ethylene and aromatic compounds), while the mole fractions of 1-alkenes from 1-butene to 1-undecene show an increase with residence time and temperature followed by a decrease. These 1-alkenes are probably primary products which decompose when temperature or residence time increase.

**Figure 5, Figure 6**

Primary products of n-dodecane thermal decomposition were determined from a selectivity analysis performed at 873K (selectivity is the ratio between the mole fraction of the considered product and the mole fractions of all products). In these conditions, the conversion of n-dodecane is less than 20% (maximum conversion corresponding to the residence time of 5s).

If the extrapolation to origin of the selectivity of a species vs. residence time is different from zero then this product is probably a primary product (Figure 7a). This is the case for hydrogen, methane, ethylene, ethane, propene and heavier 1-alkenes from 1-butene to 1-undecene from which extrapolated initial selectivities are summarized in Table 1. The sum of the initial selectivities of primary species is equal to 0.991 (theoretical value: 1). Species with the highest initial selectivities are ethylene, methane, propene and hydrogen. At 873 K, residence time has little effect on selectivities. Selectivities of hydrogen and 1-alkenes from 1-hexene to 1-undecene slightly decrease in a monotonous way with residence time whereas selectivity of ethane increases. Selectivities of 1-butene, methane and propene decrease first with residence time and then increase whereas selectivities of 1-pentene and ethylene present the opposite trend.

If the extrapolation of the selectivity vs. residence time gives a value close to zero, then the product may be a secondary product. This is the case for benzene (Figure 7b).

**Table 1, Figure 7**

**4. Modeling of n-dodecane thermal decomposition**

A first detailed kinetic model of the thermal decomposition of n-dodecane, made of 1175 reactions involving 154 species, was built using EXGAS software [14,17-19] according to rules of construction previously described [10]. Agreement between available experimental data [10] and computed results was reasonable taking into account that no kinetic parameters fitting was attempted and that only few species were quantified. Some changes were brought to this previous model: the kinetic parameters of two types of very important reactions have been modified according to data taken from literature and a sub-mechanism involving poly-aromatic compounds was added.

Comparison with available experimental data [20,21] shows that in the case of the recombination of two large radicals (containing more than two carbon atoms) the pre-exponential factors previously deduced from thermochemical data and from the A factors of the related reverse combination given by the modified collision theory [14,22] were under-estimated. Pre-exponential factor of unimolecular initiations by bond fission have been multiplied by a factor 4 (except for the unimolecular initiation involving methyl radical) in order to be in better agreement with data from the literature.

Retroene reaction is a molecular reaction which consists in a 1,5 hydrogen shift reaction followed by a dissociation [23]. Products of the reaction are a lower 1-alkene and propene (Equation 1).

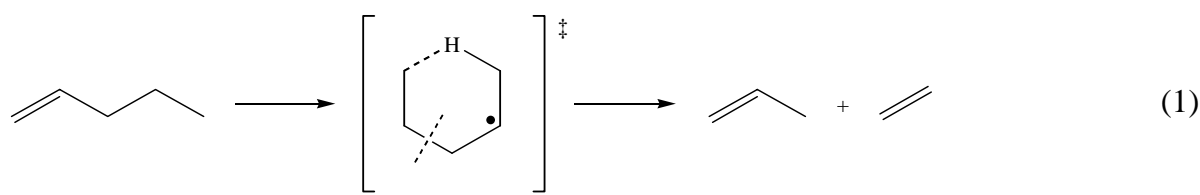

(1)

In the previous model [10], pre-exponential factor of retroene reactions, involving 1-alkenes (from 1-pentene) were taken equal to $8.0 \times 10^{12}$ cm$^3$.mol$^{-1}$.s$^{-1}$ (A factor used for the retroene reaction of decyclbenzene) [24]. Comparison with data from the literature shows that this value is slightly over-estimated in the case of 1-alkenes [25,26]. The value proposed by Tsang [25] and King [26] ($4.0\times10^{12}$ cm$^3$.mol$^{-1}$.s$^{-1}$) has been used in the model presented here.

A set of 273 elementary reactions involving 117 new species (mainly aromatics, poly-aromatics and related radical species) has been added to the modified existing model in order to explain the formation and the consumption of species such as cyclopentene, 1,3-cyclopentadiene and aromatic compounds from benzene to pyrene. These reactions have been described in two papers about propane pyrolysis [27,28]. It appears that cyclopentadiene, benzene and their related radicals (cyclopentadienyl, phenyl and benzyl radicals) play an important role in the formation of heavier aromatic and poly-aromatic species. The model explains the formation of cyclopentadienyl radical and cyclopentadiene from the addition of acetylene on propargyl radical (C$_3$H$_3$) [29]. Benzene is formed through three pathways (Figure 8): the combination of two propargyl radicals or of propargyl and allyl radicals (C$_3$ pathway) [29,30], the combination of cyclopentadienyl and methyl radicals through the formation of methyl-cyclopentadiene (C$_5$ pathway) [31-34], and the addition of acetylene to C$_4$ radicals followed by a ring closure (C4 pathway) [35].

**Figure 8**

## 5. Validation of the model of the pyrolysis of n-dodecane

Simulations were run with CHEMKIN II software [36] using the detailed kinetic mechanism presented here. Mole fractions of n-dodecane pyrolysis products have been computed as a function of the residence time for a temperature of 973 K (Figure 5) and as a function of the temperature for a residence time of 1 s (, Figure 6).

The agreement between computed and experimental mole fractions vs. residence time is globally satisfactory in the studied conditions both for conversion of n-dodecane (Figure 4a) and for the formation of hydrogen, ethylene, methane, 1,3-butadiene, indene, naphthalene and other 1-alkenes (Figure 5). Mole fraction of benzene is over-predicted for smaller residence times. Mole fractions of cyclopentene, toluene, styrene, propyne and allene are slightly under-predicted. Mole fractions of cyclopentadiene and acetylene are slightly over_predicted for short residence times and slightly under-predicted for longer residence times. Simulated conversion vs. temperature for a residence time of 1 s (Figure 4b) is in a good agreement with experimental results over a wide range of temperature (848-1073 K). Agreement between computed molar fractions and experimental data is satisfactory (, Figure 6) except for hydrogen, ethane, benzene cyclopentadiene which are slightly over-predicted and for anthracene which is under-predicted by a factor 5. The mechanism reproduces with a good accuracy the increase and the decrease of the mole fraction of 1-alkenes (from 1-pentene to 1-undecene) with increasing T.

Figure 9 displays a comparison between simulations using the new and the previous models and the experimental results obtained in a plug flow reactor between 950 and 1050 K that are previously used for validations. The model presented here leads to a higher conversion than the previous one and than experimental values, whereas it represents much better the evolution of mole fractions of main products (methane, ethylene, ethane, propene and 1-alkenes such 1-hexene) vs. residence time.

**Figure 9**

**6. Analysis of the mechanism**

Flow rate analysis has been performed for a temperature of 973 K and a residence time of 1 s corresponding to a n-dodecane conversion around 55%. As the flow rate analyses concerning the conversion of n-dodecane and the formation of the main products are similar to those previously published [10], the flow rate analysis presented in Figure 10 only shows the main pathways leading to the formation of aromatic and poly-aromatic species involved in the present mechanism. The arrows in the diagram represent the flow of consumption of the reacting species through this channel (as a percentage). This diagram underlines the importance of cyclopentadienyl, phenyl and benzyl radicals in the formation of heavier aromatic and poly-aromatic compounds. Cyclopentadienyl radical (which is mainly obtained by addition of acetylene on the propargyl radical followed by a cyclisation) is mainly consumed to form benzene through the $C_5$-pathway (Figure 8). Over-prediction of the mole fractions of benzene and cyclopentadiene vs. temperature by the model are related. Cyclopentadienyl radical is also consumed to form benzyl radical by addition to acetylene [37] and to produce naphthalene by combination with itself followed by several rearrangements [38]. Phenyl radical mainly reacts by metatheses to give back benzene. It is also consumed through minor pathways to form styrene by addition to ethylene, phenyl-acetylene by addition to acetylene and biphenyl by addition to benzene. Benzyl radical is consumed to form toluene by metathesis and combination with hydrogen radical, ethyl-benzene by combination with methyl radical and indene by addition of acetylene. Indene reacts by metatheses to lead to indenyl radical which is then consumed to form phenanthrene by combination with cyclopentadienyl radical and methyl-indene by combination with methyl radical. Methyl-indene reacts by metatheses to form methyl-indenyl radical, which leads to naphthalene by an intramolecular rearrangement. Naphthalene is consumed through metatheses to give radicals leading to acenaphthalene (addition of acetylene) and methyl-naphtalene (combination with methyl radical). Phenanthrene leads to pyrene (through metatheses and addition on acetylene) and to anthracene.

**Figure 10**

Sensitivity analyses related to the mole fractions of n-dodecane, 1-hexene and benzene have been performed at $\tau=1$s and at three temperatures (873, 973 and 1073 K). They show that the conversion of n-dodecane is mainly controlled by unimolecular initiations and by H-abstractions from reactant with methyl radicals at low temperature and by H-abstractions from reactant with hydrogen atoms at high temperature (Figure 11a). Combination of two methyl radicals is important at low temperature because the reaction is competitive with metatheses involving methyl radicals.

The formation of long 1-alkenes is mainly ruled by the reaction of decomposition by β-scission of linear dodecyl radicals obtained from metatheses of radicals with n-dodecane. Two types of reactions have a great influence on their consumption as shown in Figure 11b for 1-hexene: the decomposition by retroene reaction and the decomposition by unimolecular initiation reaction. Unimolecular initiation (energy of activation of 70.7 kcal.mol$^{-1}$) becomes more important compared to retroene (energy of activation of 57.5 kcal.mol$^{-1}$) when temperature increases.

As far as benzene is concerned sensitivity analysis shows that it is very influenced by the reaction leading from the methyl-cyclopentadiene to the methyl-cyclopentadienyl radical by H-abstraction with hydrogen and methyl radicals at low and high temperatures. At low temperature (873 K) unimolecular initiations of n-dodecane, the reaction of combination of two methyl radicals (this reaction is competitive with the reaction of combination of methyl and cyclopentadienyl radical) and the reactions of metathesis involving ethyl radical and n-dodecane (this reaction forms ethane which indirectly leads to intermediate species in the formation of benzene) have a great influence on benzene formation (Figure 11c).

**Figure 11**

## 7. Conclusion

The thermal decomposition of n-dodecane has been studied in a jet-stirred reactor over a wide range of temperatures from 793 K to 1073 K and for residence times between 1 and 5 s. 32 products of the reaction have been analyzed. Major products are hydrogen and ethylene. Other products are mainly 1-alkenes from propene to 1-undecene and aromatic compounds. The formation of aromatic compounds intervenes rather quickly; it takes importance for conversions higher than 25 %. The formation of large amount of aromatic and polyaromatic compounds may me troublesome: these species are precursors of the formation of soot and even coke at high temperature [39Mendiara et al., 2005] which can generate a solid deposit in the cooling channel (before the injection in the combustion chamber) and a fouling in the combustion chamber.

A detailed kinetic model generated by the EXGAS software has been completed by a set of elementary reactions involving aromatic and poly-aromatic compounds. The new model obtained is made of 1449 reactions and contains 271 species. It is expected that this detailed kinetic model (composed of elementary steps) will be accurate enough for temperatures from 500 to about 1500 K.

Agreement between new experimental results presented here and computed results is satisfactory especially for primary products such 1-alkenes. As far aromatic compounds are concerned a light variation between experiments and modeling is observed especially for long residence times. This may be related to uncertainties on kinetic parameters of reactions involved in the formation of benzene (C5-pathway) and poly-aromatics compounds. The new model reproduces well experimental data presented in a previous paper [10] and obtained with a different type of reactor working in different conditions. Simulations of the cooling a engine will be performed from the model of the thermal decomposition of n-dodecane which will be coupled with computational fluid dynamics codes. The size of the detailed kinetic model (hundreds of species and thousands of reactions) involves very long computational times. Models can be reduced and adapted to have lower calculation times. The lumping is one of the methods which can be used for the reduction of detailed kinetic models. It consists in gathering molecules which have the same formula, present identical functional groups and include rings of the same size into one species. This method was used in the secondary mechanism of the present model and applied to linear alkanes and dienes which are formed in small quantities [10]. A second way of reducing models is to eliminate redundant species and reactions from a kinetic analysis (rate of production and rate sensitivity analyses) of the mechanism. In some cases this method allows to eliminate one third of the species and two thirds of the reactions [40]. Reduction based on the investigation of time-scales consists in identifying QSSA species and then to eliminate them by solving the QSSA expressions or by writing global reactions [40]. At least another method for the reduction of the kinetic models is the mathematical fitting (by a system of ordinary differential equations for example) of simulated results obtained with the detailed model [40].

**Acknowledgments**

This work was supported by MBDA-France and the CNRS. The authors are very grateful to E. Daniau, M. Bouchez and F. Falempin for helpful discussions.

**Figure captions:**

Figure 1 : Experimental apparatus flow sheet.

Figure 2 : Scheme of the jet stirred reactor and the annular preheating.

Figure 3 : Heavy products separation (HP-1; FID).

Figure 4 : Conversion of n-dodecane versus (a) residence time and (b) temperature ($\tau=1s$).

Figure 5 : Evolution of products mole fraction with residence time (T=973 K).

, Figure 6 : Evolution of products mole fractions with temperature ($\tau=1s$).

, Figure **7** : Selectivity of (a) ethylene, (a) 1-hexene and (b) benzene vs. residence time (T=873 K).

Figure 8 : The three pathways to benzene formation.

Figure 9 : Comparison between previous experimental data (plots) and modeling using the new model (solid line) and the previous model (dashed line) (plug flow reactor, T=1050 K).

Figure 10 : Aromatics and poly-aromatics consumption flux analysis (T=973 K; $\tau=1s$)

Figure 11 : Sensitivity analyses for (a) n-dodecane, (b) 1-hexene and (c) benzene ($\tau=1s$).

Table 1

Initial selectivities of primary products obtained by extrapolation (T=873 K).

| Primary product | Initial selectivity |
|---|---|
| hydrogen | 0.101 |
| methane | 0.135 |
| ethylene | 0.322 |
| ethane | 0.030 |
| propene | 0.124 |
| 1-butene | 0.065 |
| 1-pentene | 0.040 |
| 1-hexene | 0.050 |
| 1-heptene | 0.035 |
| 1-octene | 0.030 |
| 1-nonene | 0.026 |
| 1-decene | 0.024 |
| 1-undecene | 0.009 |
| *total* | *0.991* |

Figure 1

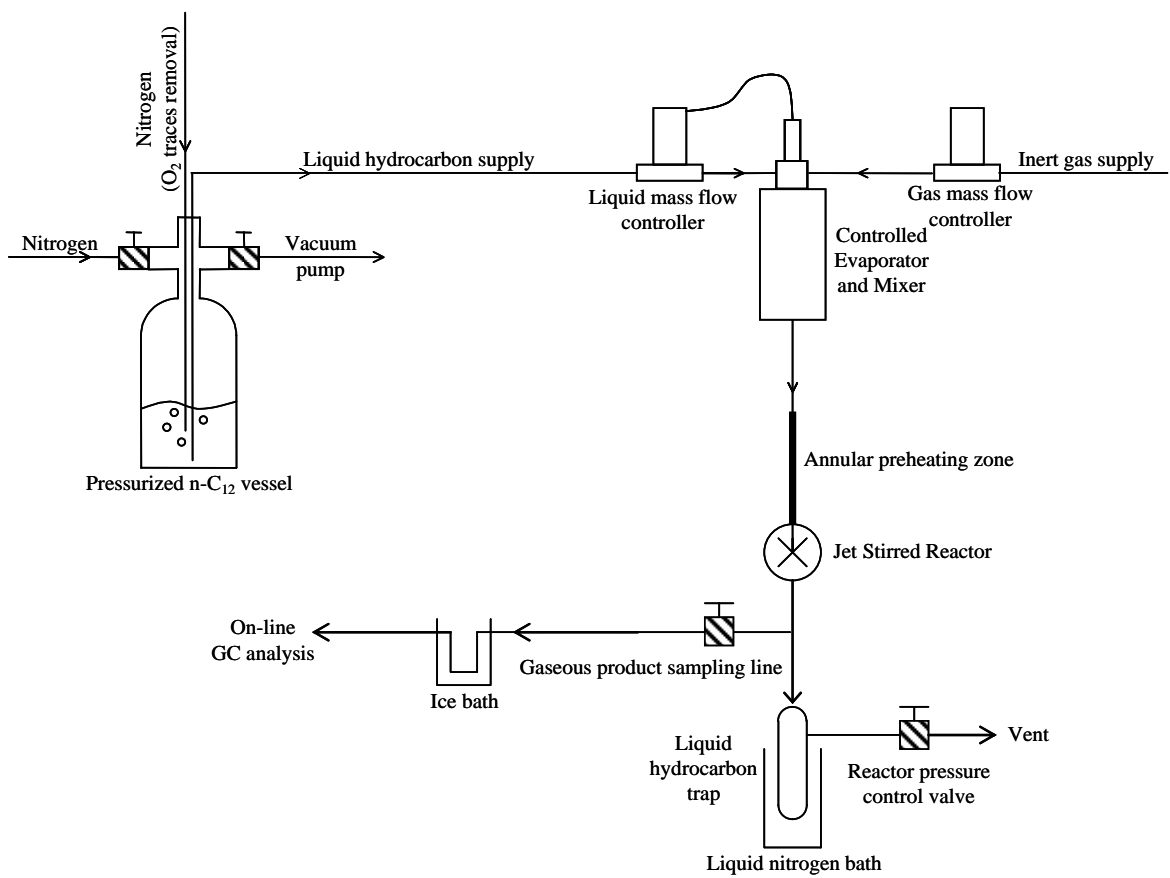

Figure 2

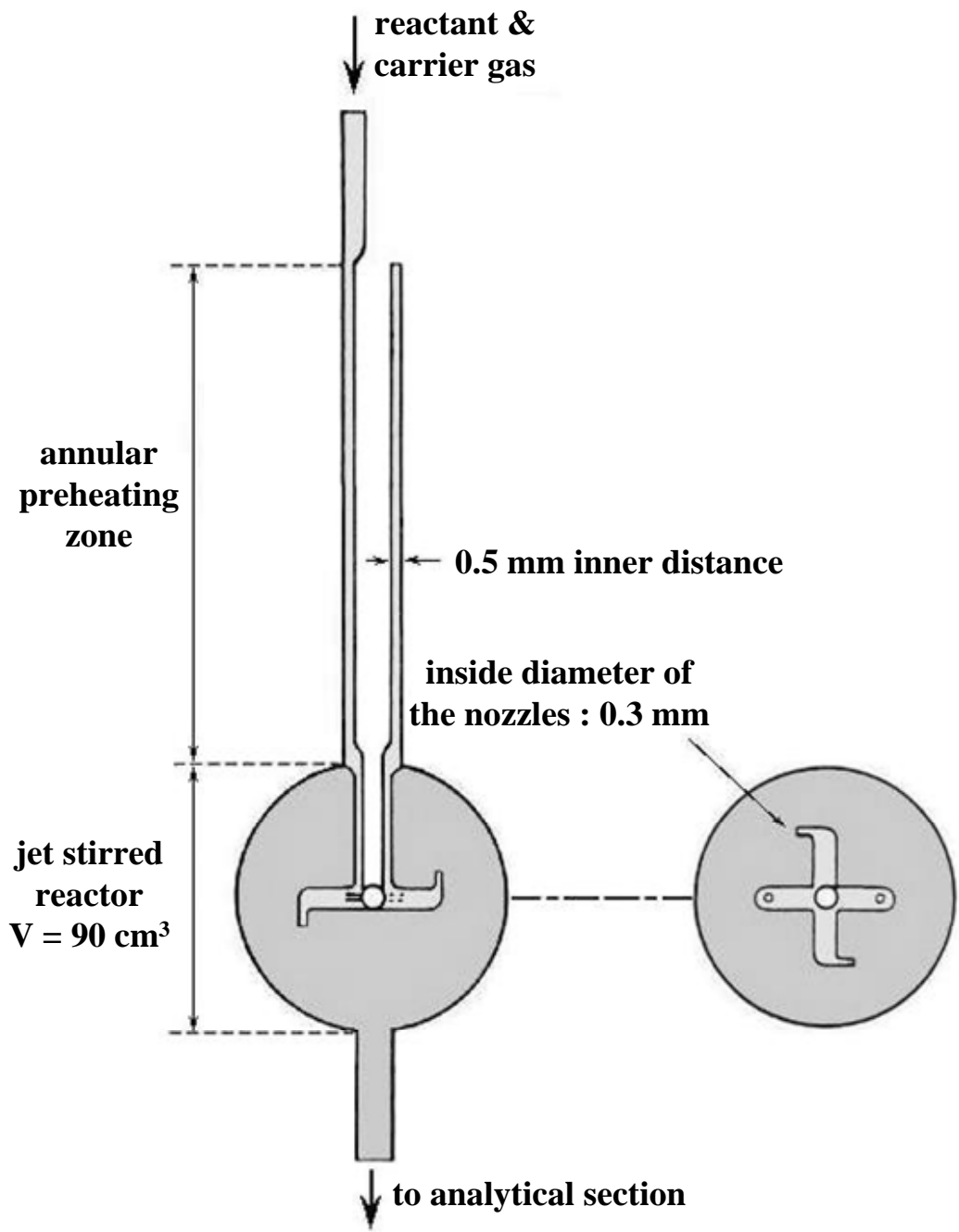

Figure 3

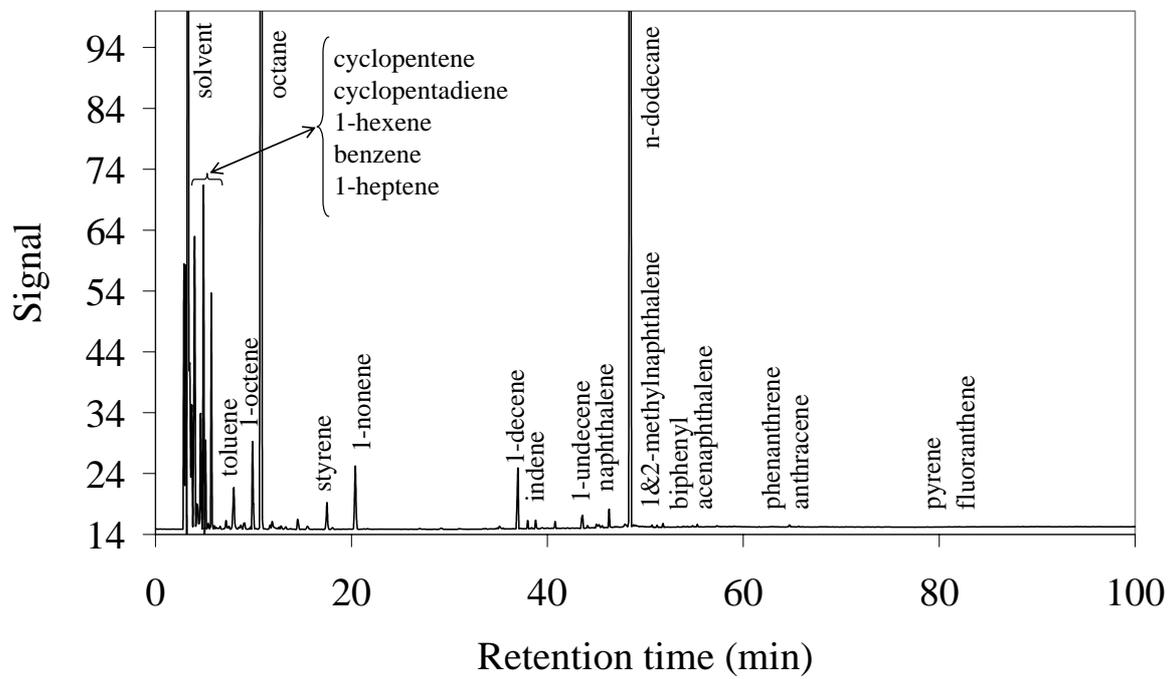

Figure 4

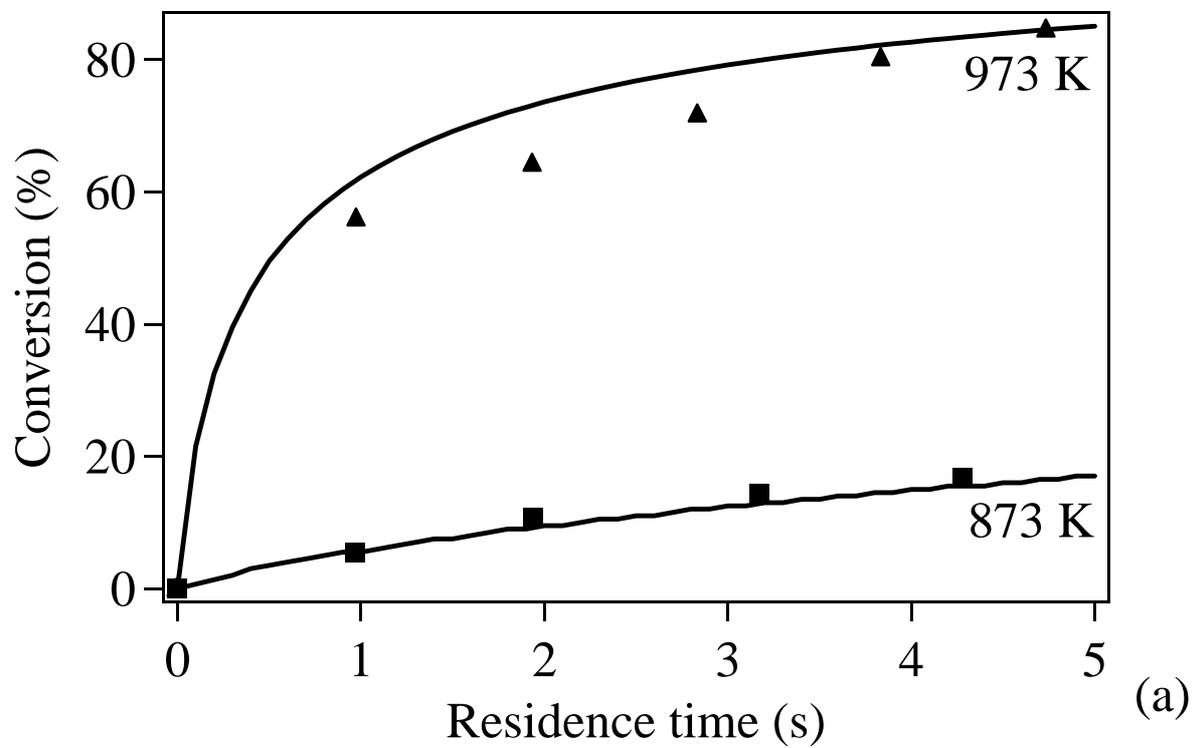

(a)

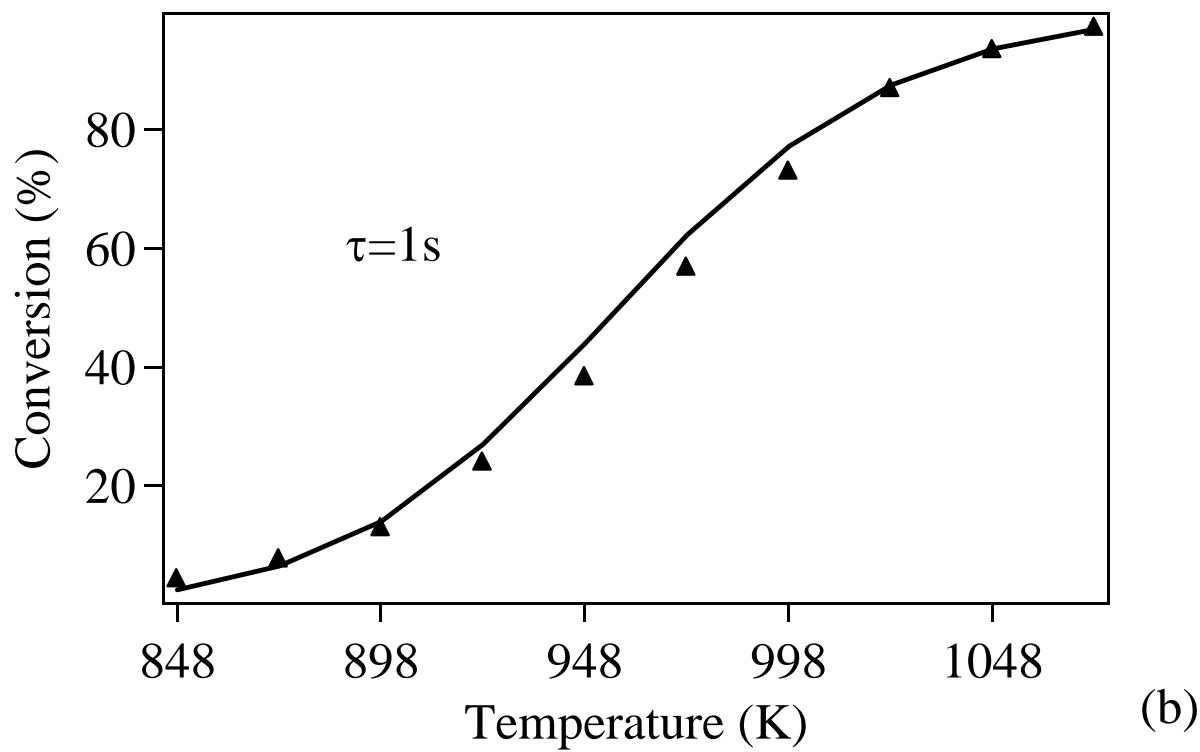

(b)

Figure 5

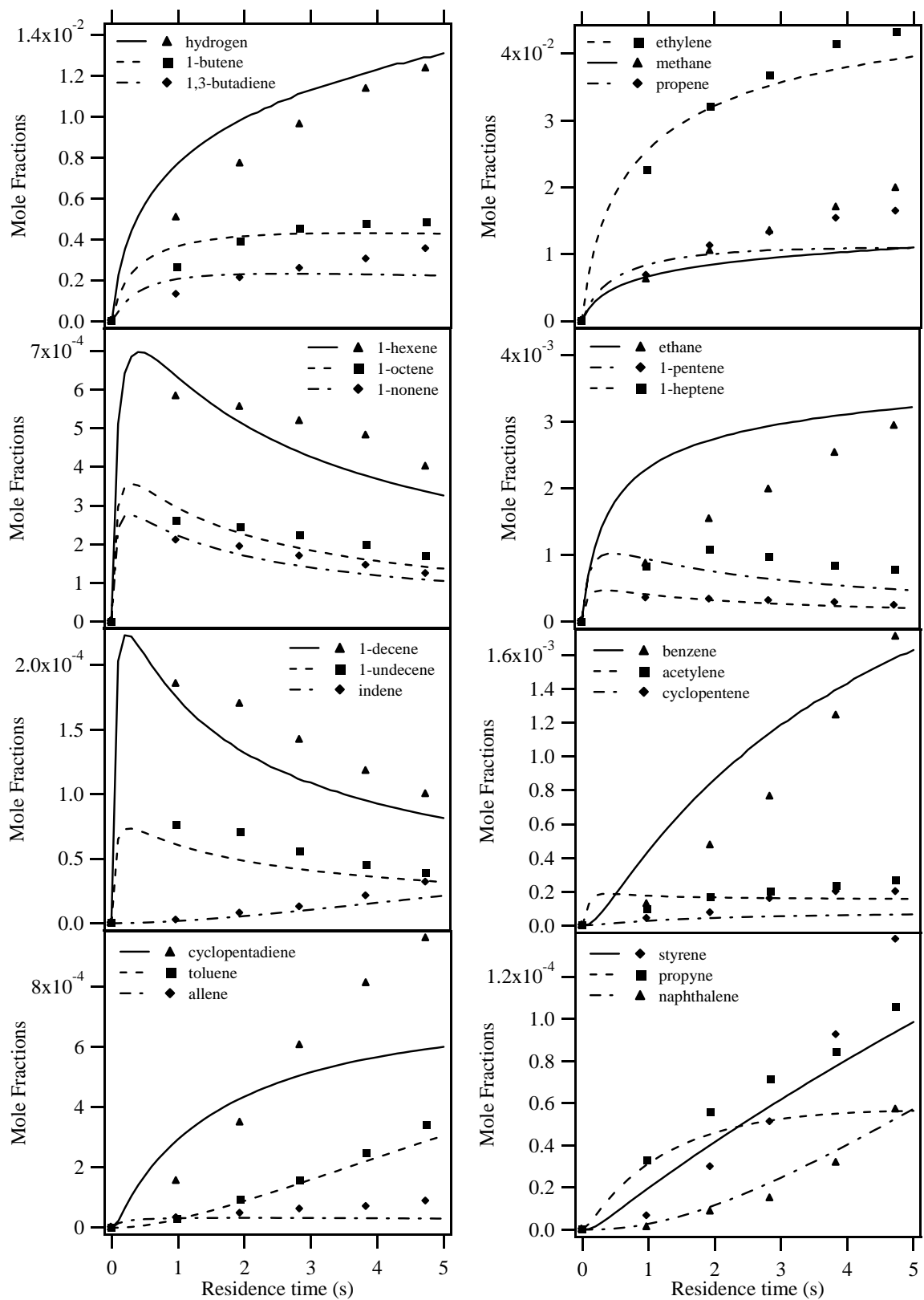

, Figure 6

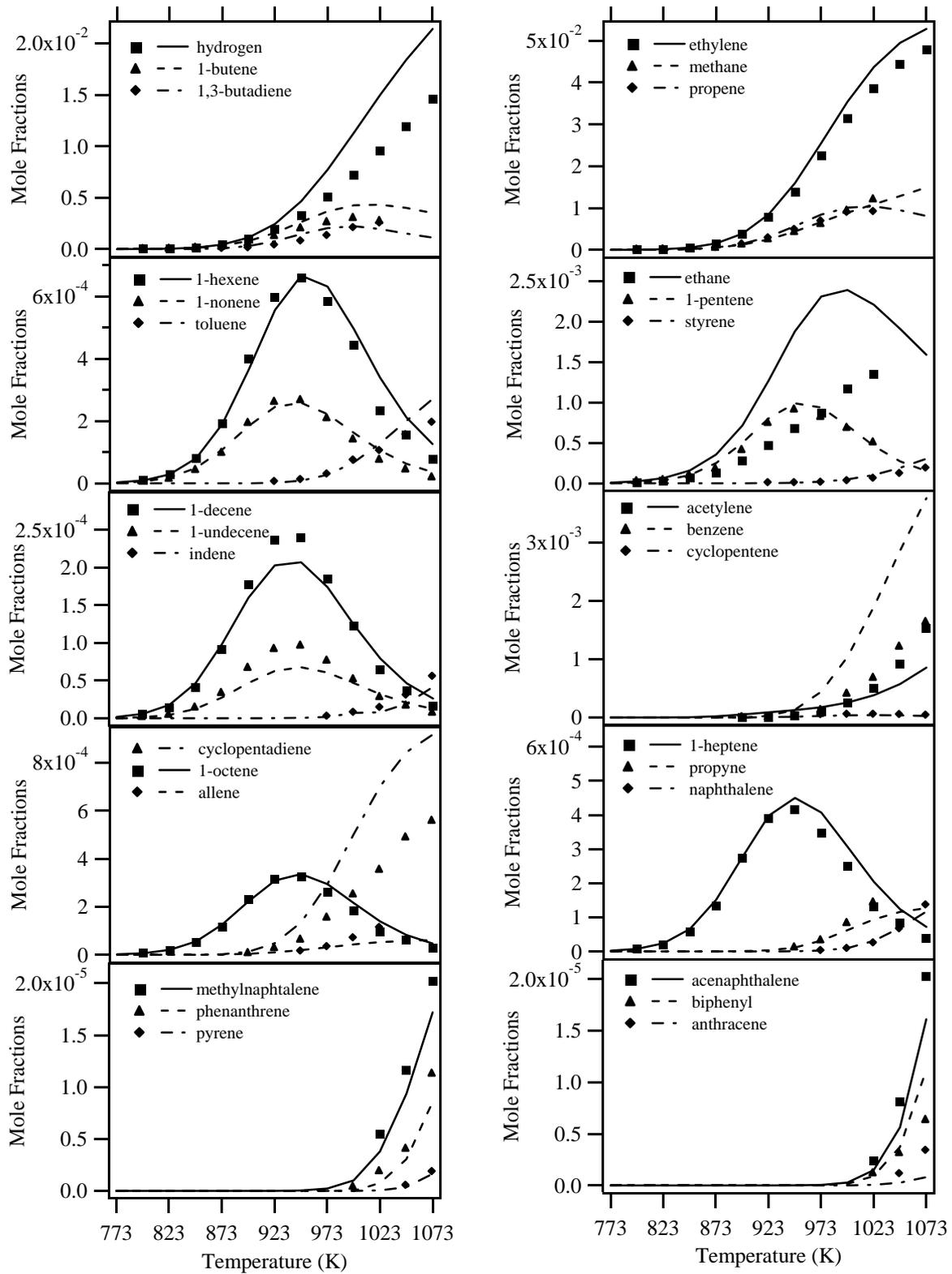

, Figure 7

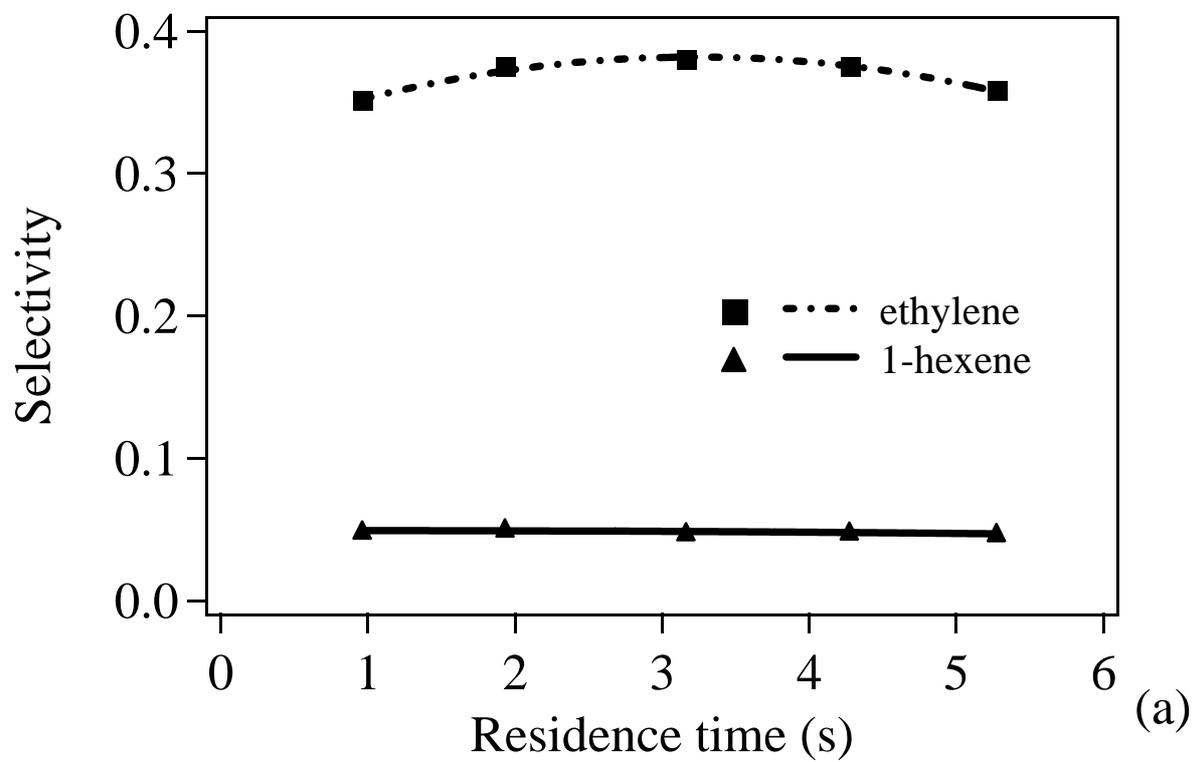

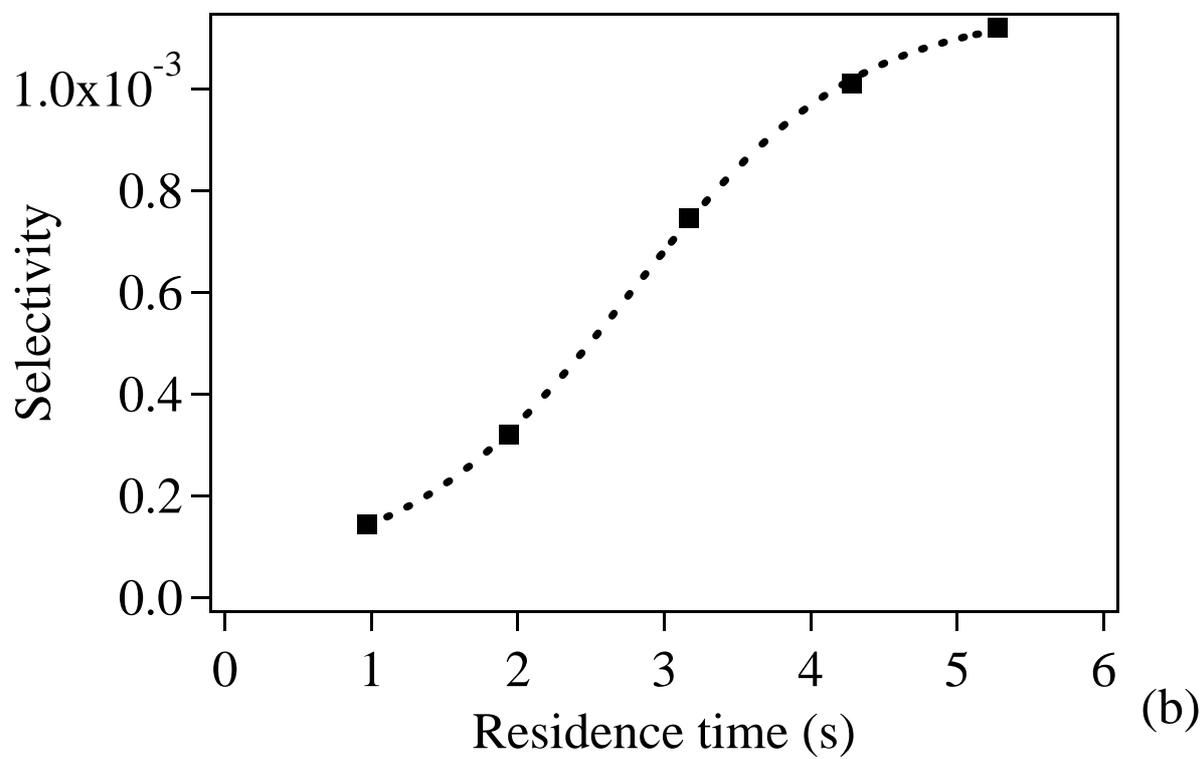

Figure 8

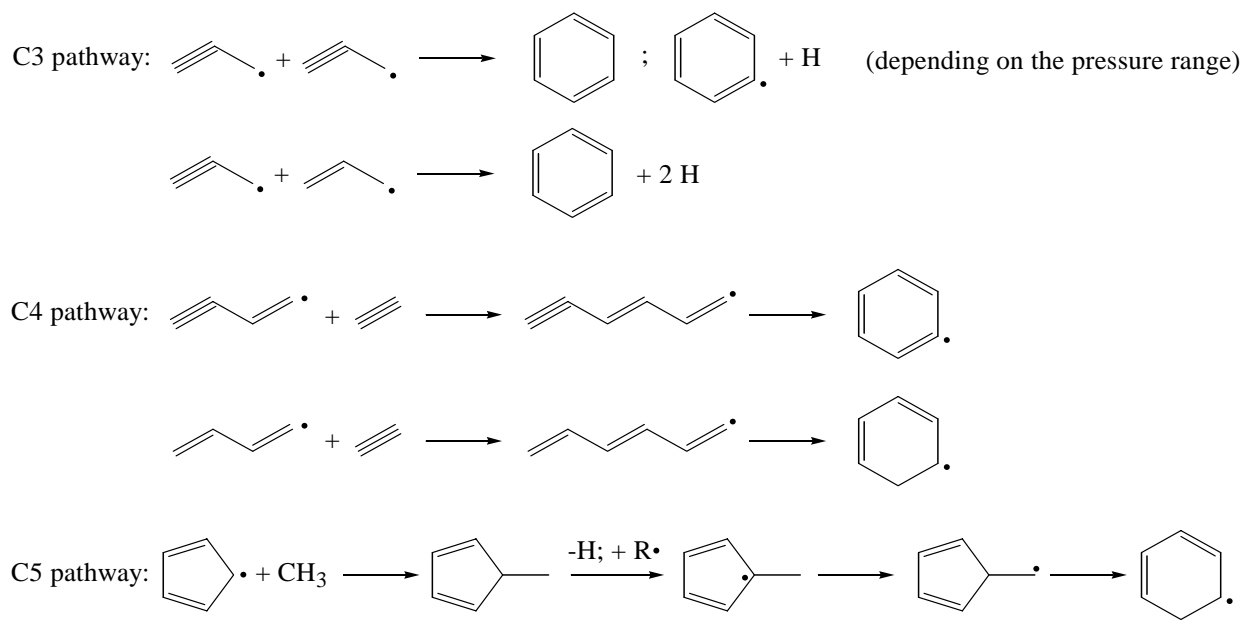

Figure 9

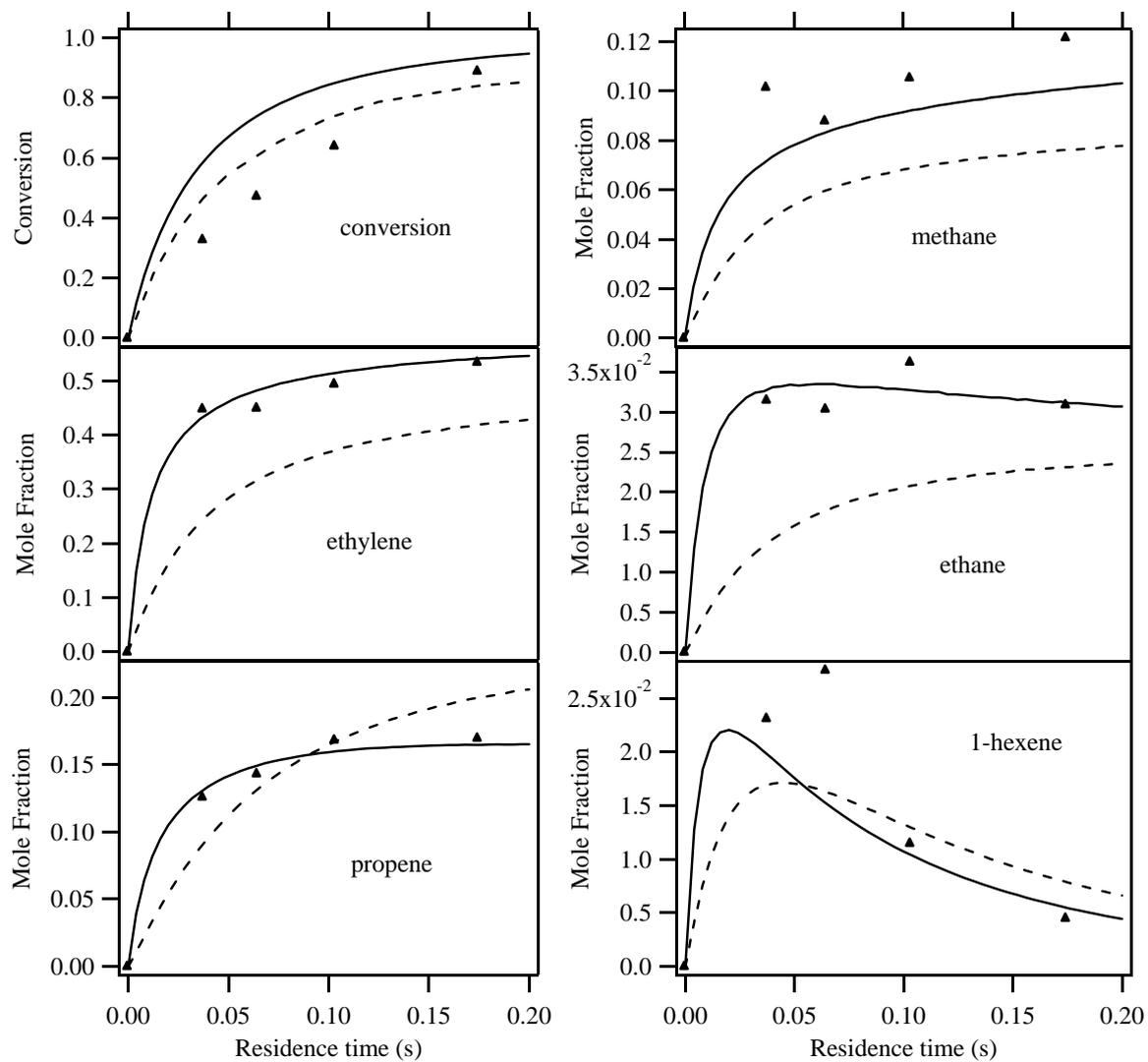

Figure 10

Figure 11

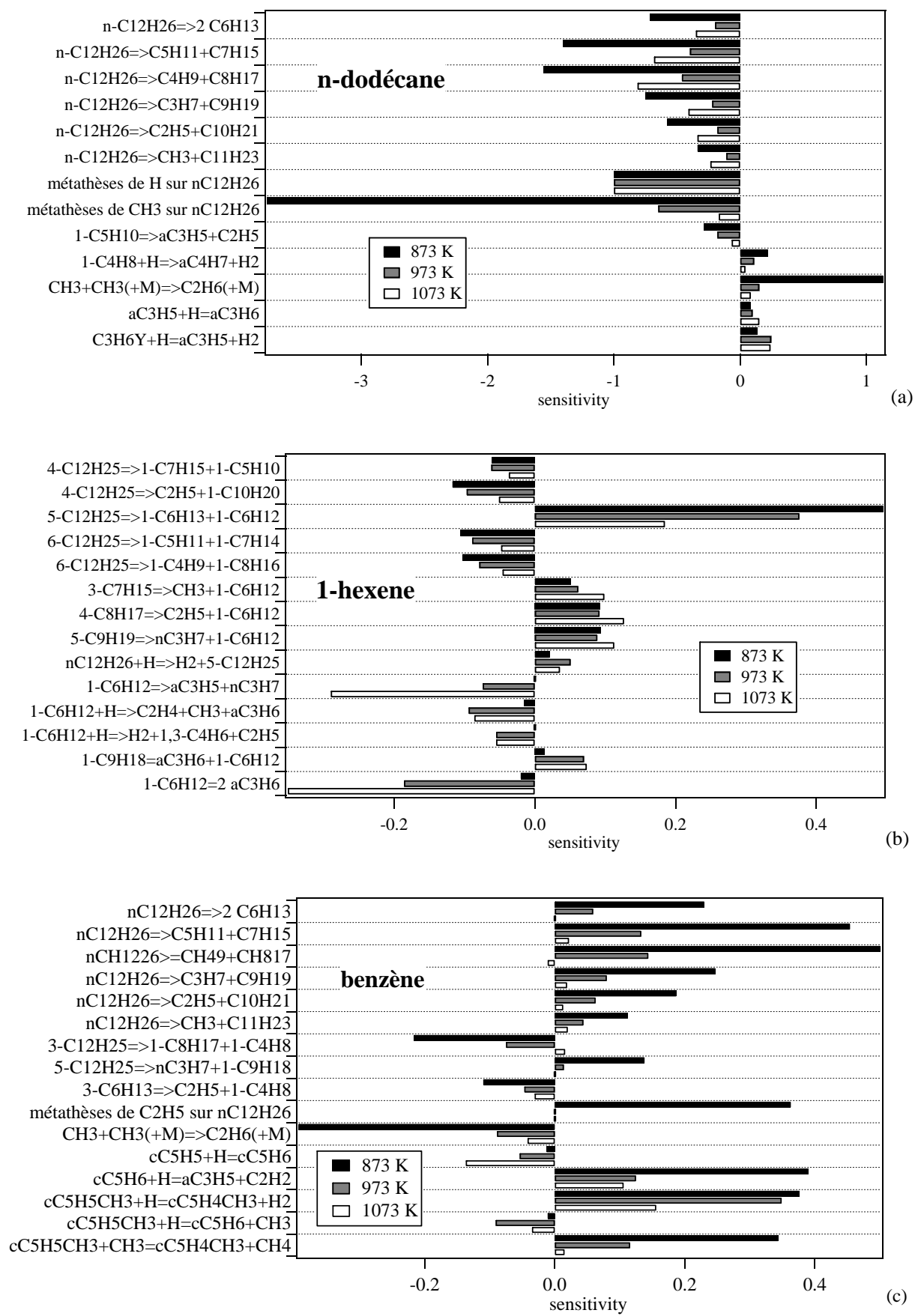